\documentclass[12pt,preprint]{aastex}

\def\omega0{\Omega_{\rm m,0}}
\def\lambda0{\Omega_{\Lambda,0}}

\def\fsub{f_{\rm sub}}
\def\fcool{f_{\rm cool}}
\def\rv{r_{\rm vir}}
\def\reff{r_{\rm eff}}
\def\kms{\,\rm km\,{s}^{-1}}

\def\mpc{\,\rm Mpc}
\def\LCDM{\Lambda{\rm CDM}}

\def\msun{M_\odot}
\def\pc{\,{\rm pc}}

\def\beq{\begin{equation}}
\def\eeq{\end{equation}}

\begin{document}

\title{Anomalous Flux Ratios in Gravitational Lenses: For or Against 
CDM?}
\author{Shude Mao\altaffilmark{1},
Yipeng Jing\altaffilmark{2}, Jeremiah P. Ostriker\altaffilmark{3},
Jochen Weller\altaffilmark{3}
}
\altaffiltext{1}{ University of Manchester, Jodrell Bank Observatory, 
  Macclesfield, Cheshire SK11 9DL, UK; smao@jb.man.ac.uk}
\altaffiltext{2}{ Shanghai Astronomical Observatory; the Partner Group
of MPA, Nandan Road 80, Shanghai 200030, China; ypjing@center.shao.ac.cn}
\altaffiltext{3}{ Institute of Astronomy, Cambridge University,
Madingley Road, Cambridge, UK; (jpo, jw249)@ast.cam.ac.uk}

\shorttitle{Lensing and substructures in CDM}
\shortauthors{Mao, Jing, Weller \& Ostriker}

\begin{abstract}
We review the evidence for substructures from the anomalous flux ratios
in gravitational lenses. Using high-resolution numerical simulations, 
we show that at typical image positions, the fraction of surface mass
density in substructures is $\la 0.5\%$ with mass above
$10^{-4}$ virial masses in the ``concordance''
$\LCDM$ cosmology. Substructures outside
the virial radius (but projected at typical lens image positions)
only increase the fraction moderately. Several effects, in particular
baryonic settling and the requirement of compactness, 
may further decrease the predictions by a factor of few.
The predicted fraction with appropriate properties 
thus appears to be lower than that required 
by lensing, although both are still uncertain.
More speculative substructures such as massive black holes 
($M \sim 10^5-10^6 M_\odot$) in the halo may offer viable alternatives.
\end{abstract}


\section{INTRODUCTION}

Gravitational lenses on arcsecond scales
provide a unique sample to probe the mass distribution
in the lensing galaxies at intermediate redshift ($z\sim 0.5-1$).
Image positions in most lenses can be fitted adequately
using simple smooth galaxy mass models. But
observed flux ratios are  more difficult to match
(e.g., Kochanek 1991). The discrepancy
between the predicted and observed flux ratios is
commonly referred to as the ``anomalous flux ratio problem.''
The most apparent cases are found in
quadrupole lenses where we observe a close pair or a close triple of images.
Here we know that the lensed source is close to 
either a fold or a cusp caustic. The {\it asymptotic} magnification behavior 
in such cases is well understood -- a close pair must have equal
brightness, while
for a close triple, the flux of the middle image should be equal to the
total fluxes of the two outer images. Virtually all the observed
pairs and triples disagree with these relations (\S2).
This has been argued as evidence for substructures 
on the scale of the separations of the images
(a few tenths of arcseconds, e.g., Mao \& Schneider 1998;
Metcalf \& Zhao 2002). Another piece of evidence
for substructures is the
fact that saddle images are preferentially
dimmed compared to model predictions (Kochanek \& Dalal 2003).
This is expected from milli-lensing by substructures (Keeton 2003) or microlensing
by stars (Schechter \& Wambsganss 2002)\footnote{We refer to
lensing by substructures and lensing by stars as milli-lensing and
microlensing respectively because the angular scales
involved are $\sim$ mas and $\mu$as in these two cases.};
such a preferential
de-magnification of saddle images is not expected from 
other propagational effects.

The Cold Dark Matter (CDM) structure formation model predicts
the existence of just such substructures
from both semi-analytical studies and numerical simulations
(e.g., Kauffmann et al. 1993; Klypin et al. 1999;
Moore et al. 1999; Ghigna et al. 2000). About 5-10\% of 
the mass is predicted to be in substructures with a 
mass spectrum of $n(M)dM \sim M^{-1.8}dM$. Intriguingly, the predicted  number of
subhaloes in CDM exceeds the observed number of {\it luminous}
satellite galaxies in a Milky-Way type galaxy 
(e.g., Klypin et al. 1999; Moore et al. 1999; see also Stoehr et al. 2002).
One solution for this disagreement may be that some 
substructures, especially those of lowest mass, are dark. If this is true, then gravitational lensing
may be the best way to detect them. In this paper, we 
examine the required amount of substructures in gravitational lenses
(\S2) and compare it with the predictions from
CDM (\S3). Finally, in \S4, we discuss several effects that affect the 
predictions of numerical simulations. 
Throughout this paper, we adopt the ``concordance'' $\LCDM$ 
cosmology (e.g., Ostriker \&
Steinhardt 1995; Spergel et al. 2003 and references therein),
with a density parameter $\omega0=0.3$, a cosmologically constant
$\lambda0=0.7$, a baryon density parameter $\Omega_{\rm b}=0.024h^{-2}$,
and we take the power-spectrum normalization $\sigma_8=0.9$. We write the
Hubble constant as $h=H_0/(100\kms\mpc^{-1})$ with $h=0.7$.

\section{Required Amount of Substructures From Gravitational Lenses}

A number of papers have discussed evidence for substructures
from gravitational lenses; most concentrated on the anomalous flux
ratios (e.g., Mao \& Schneider 1998; 
Metcalf \& Madau 2001; Chiba 2002; Bradac et al. 2002; Dalal \&
Kochanek 2003; Keeton 2003), while several papers discussed
astrometric signatures, such as bent jets in B1152+199
(Metcalf 2002), and unusual VLBI structures  for MG2016+112
(Kochanek \& Dalal 2002; see also Koopmans et al. 2002)
and B0128+437 (Phillips et al. 2000; Biggs et al. 2003).
It is illustrative
to see the issues using the largest homogeneous lens survey --
the Cosmic Lens All-Sky Survey (CLASS, Myers et al. 2003 and
Browne et al. 2003). This radio survey has well defined selection criteria and
does not suffer from the effect of dust extinction. Radio lenses are
also not substantially affected by stellar microlensing 
(see Koopmans et al. 2003 and references therein).

In total, there are 22 new lenses discovered in CLASS.
Among these, 7 are simple quadrupole lenses, 
including 5 close pairs
(B0128+437, Phillips et al. 2000;
MG0414+0534, Hewitt et al. 1992;
B0712+472, Jackson et al. 1998;
B1608+656, Myers et al. 1995;
B1555+375, Marlow et al. 1999),
and two close triples
(B2045+265, Fassnacht et al. 1999; 
B1422+231, Patnaik et al. 1992).
For the five pairs listed above, the observed flux ratios are
$0.56$ (5GHz), $0.88$ (15GHz), $0.75$ (15GHz),
$0.51$ (8.4GHz), $0.56$(15GHz);
for each lens, the highest frequency where the pair has been
observed is shown in the bracket. For the
two close triples, the ratios of the flux of the middle image
to the total flux of the two outer images are, $0.32$
(14.9GHz) and $0.70$ (15GHz, Patnaik \& Narasimha 2001)
respectively. The typical errors on the flux ratios are $\sim 1-2\%$.
Except the case of MG0414+0534, the observed
values are all different from the asymptotic value (unity).

However, not all these flux ratios are ``anomalous'' because some of
these systems may not yet reach the {\it asymptotic} regime. 
Also some radio lenses may have been affected by scattering by free electrons
along the line of sight. In addition, most lens modelers adopt isothermal ellipsoidal
models or variants as an approximation for the lensing potentials.
Models with more complex radial and angular structures can usually  better
reproduce the observed flux ratios (e.g., Evans \& Witt 2003; however, see
Kochanek \& Dalal 2003). Observationally, it is clear that some lens
systems are complex. For example, B1608+656 has two lensing galaxies and a model
which accounts for this fact can reproduce the observed flux ratio
(Koopmans \& Fassnacht 1999). For B0712+472, a singular isothermal
ellipsoid (SIE) model cannot reproduce the flux ratio, but a foreground group of galaxies
was subsequently found along the line of sight of the primary lensing galaxy
(Fassnacht \& Lubin 2002), so a more realistic model may match the flux ratio.
The close pair flux ratio in B1555+575 cannot be reproduced
by an SIE model (Meyers et al. 1995), but a more complex
model (e.g., with an additional shear) may be able to explain it.
For B0128+437, an SIE model
cannot fit the flux ratio, but the observed flux ratio can be 
reproduced with an additional shear. However, in this case,
the relative orientations of jets in this system
appear difficult to match with a smooth model (Biggs et al. 2003),
so substructures may yet be called for. For 
the two close triple lenses, the flux ratios appear
difficult to match with smooth models (Mao \& Schneider 1998; Fassnacht
et al. 1999).

Mao \& Schneider (1998) showed that substructures preferentially
affect the flux ratios of highly magnified images. In order to 
reproduce the flux ratios in B1422+231, they found that 
the required perturbation in the dimensionless surface mass
density, $\delta\kappa/\kappa$, is of a few percent and
roughly corresponds to a physical surface density of a few tens
$\msun\pc^{-2}$. Dalal \& Kochanek (2002) performed a statistical study
of seven radio lenses using Monte Carlo simulations, six of which show
anomalous flux ratios. They find that the fraction of mass required
in substructures should be in the range of $\fsub =0.6\%-7\%$ (90\% 
confidence limit) with
a best fit of $\fsub \approx 2\%$. More recently, Metcalf et al. (2003)
applied the method of Moustakas \& Metcalf (2003) to the quadrupole system
2237+0305. They conclude that, in order to match the observed flux ratios
in the radio, infrared and narrow and broad emission lines, 
4\%-7\% of the surface mass density (95\% confidence limit)
must be in substructures with mass between $10^4M_\odot-10^8 M_\odot$. 

\section{Predicted Substructures in Numerical Simulations}

We use the high resolution halo simulations of Jing \& Suto (2002;
2000) to constrain the fraction of mass in substructures; similar
results (but with larger error bars) are found using the
simulation data obtained with the tree-particle-mesh code of Bode \& Ostriker (2003).
Twelve halos were selected from a cosmological simulation of box of $100h^{-1}\mpc$, with
four each at galactic, group, and cluster masses, respectively. These
simulations are evolved with a nested-grid PPPM code which was
designed to simulate high-resolution halos.  The force resolution is
typically $0.004 \rv$, where $\rv$ is the virial radius.  
At the end of each simulation, about $(0.5 - 1)\times 10^6$
particles are within $\rv$ of each halo (see
Jing \& Suto 2000 for details).

We adopt the {\tt SUBFIND} routine of Springel et al. (2001) to find
disjoint self-bound subhalos within a parent halo. All subhalos with
more than 10 particles are included in our analysis.
Our numerical resolution implies that we can only identify subhaloes
with mass larger than about $10^{-4}$ of the parent halo mass;
the most massive subhalo has about $10\%$ of the parent halo
mass. For each halo, we make 30 random projections and calculate the
total mass within different annuli; the annuli 
are equally spaced in $\log R/\rv$ from $-2.2$ to 0 with
a step size of 0.2, where $R$ is the projected radius and
the spherical radius is $r$.
The total mass in substructures within an annulus is calculated by summing
up all the mass of subhalos whose centers fall in it. Dividing the total
substructure mass by the total mass within the annulus yields
the fraction of the projected surface density in substructures, $\fsub$. The mean
value and variance of $\fsub$ are found from the 12 haloes and the 30 random projections.
In the upper left panel of Fig. 1, we show $\fsub$
as a function of $R/\rv$. The mean 
fraction can be approximated as $\fsub \approx 0.25 (R/\rv)$. The scatter around the mean
among haloes is quite large. At $R \approx \rv$, the scatter is about 40\%; 
at smaller $R$ the scatter is larger due to fewer subhaloes along the line of sight.
As lensing concerns only the projected surface density, we also
checked whether substructures outside the spherical virial radius can contribute 
to the surface density at a given $R$ (Fig. 1, top right).
The increase is about a factor of two at $R \approx \rv$ (cf. Klypin et
al. 2001), but more modest ($\la 1.3$)
at typical image positions ($R \sim 0.01\rv-0.03\rv$), consistent with the
analytical estimate by Chen et al. (2003). 

Almost all subhaloes in the simulation have a spherical radius $r>0.1\rv$.
Most substructures at typical image positions ($R \sim 0.01\rv-0.03\rv$)
only appear along the line of sight due to projection. These subhaloes, especially those at $r
\sim \rv$, are extended, thus not all the bound
mass within the subhaloes can efficiently cause the flux anomaly (see \S4).
To illustrate this effect, $\fsub$ is recalculated by
including only the mass within a spherical radius of
$0.025\rv$ for each subhalo. Due to this compactness requirement, at 
$R = 0.03\rv$, the average $\fsub$ value (for substructures with
$r<\rv$) is lowered from 0.8\% to 0.6\% (see the lower panels in Fig. 1).
We return to this compactness issue in the discussion.

To compare the predicted fraction with lensing requirement we
need an estimate of the virial radius for the lensing
galaxies. If the lens and
source redshifts are known, then the separation
between images ($\propto \sigma^2$) can be used to determine
the velocity dispersion ($\sigma$) and the approximate 
halo circular velocity ($V_{\rm c} \approx \sqrt{2} \sigma$), which
in turn allows us to determine the halo
virial radius (e.g., eqs. 2-3 in Mo, Mao \& White 1998). The lens
and source redshifts are known for 
MG0414+0534, B0712+472, B1608+656, 
B2045+265 and B1555+375. For these
systems, the image positions in units of the 
virial radius are, 0.027, 0.012, 0.018, 0.017 and
0.012, respectively. From Fig. 1 we can then infer the fraction of
mass in substructures is only around 0.2-0.8\%, with a large
scatter among different haloes.

So far we have only considered the fraction of {\it dark matter} surface
density in substructures. However, the inner parts of galaxies are likely
dominated by baryons. To address this issue, we consider the cooling and
settling of baryons in an NFW halo (Navarro, Frenk \& White 1997), similar
to the procedure used by Keeton (2001). Initially
the baryons and dark matter follow the same
profile. The baryons then cool and condense into the center to form
a stellar component with a de Vaucouleurs profile while
the dark matter responds to the baryonic settling adiabatically.
The model is described by the
concentration parameter $c$ in the NFW profile, the
effective radius ($\reff$) in units of the halo virial radius,
and the ratio of the stellar mass to the total
mass, $\fcool$ (or equivalently, a mass-to-light ratio).

For the five sources with known lens and source redshifts (MG0414+0534, 
B0712+472, B1608+656, B2045+265 and B1555+375),
the effective radii have been determined using HST photometry
(Kochanek et al. 2000). In units of
the virial radius they are, respectively, 0.018, 0.006, 0.01,
0.006 and 0.005. We take $c=10$ (appropriate for a galactic-sized halo)
and then carry out the procedure described above.
For the five systems, we find that 
the stellar component contributes about 20\%-50\% of the projected
surface density at the image positions ($R \approx 1.5\reff-3\reff$)
for $\fcool$ ranging from 0.05 to 0.16 
($\approx \Omega_{\rm b}/\Omega_0$).
The fraction of surface mass density in dark matter is consistent with
that required by stellar microlensing ($\sim 70-90\%$ at 1.5$\reff$, Schechter
\& Wambsganss 2003),
but at odds with the recent claim of Romanowsky et al. (2003).
The effect of baryons hence reduces the 
the mass fraction in substructures in typical lensing systems
to $\la 0.5\%$. 

\section{Discussion}

  We have reviewed the evidence for substructures from close pairs and
triples in quadruple lenses. As emphasized by
Kochanek \& Dalal (2003), the fact that saddle images are frequently dimmer than expected
is difficult to accommodate by other means. Quantitatively,
the anomalous flux ratios in lenses appear to require a few percent of 
the surface mass density in substructures at typical image positions
(Dalal \& Kochanek 2002; Metcalf et al. 2003).
The required fraction is higher than that provided
by globular clusters and luminous satellite galaxies (Mao
\& Schneider 1998; Chiba 2002) and it also
appears to be higher than the predicted values 
($\fsub \la 0.5\%$) from the $\LCDM$ cosmology at typical image positions.
However, at present it is unclear how serious the discrepancy is because
of uncertainties in both observations and theoretical predictions. 

There are a number of issues that need to be understood better
in current numerical simulations. Even the basic question of the
identification of substructures  needs to be
explored further. The {\tt SUBFIND} algorithm we adopted only identifies bound subhaloes,
however, tidal streams from disrupted systems (for examples in the Milky Way, see, e.g.,
Ibata et al. 2001; Yanny et al. 2003) may also contribute to 
the budget of substructures. Another important issue is whether our
results have achieved  convergence  as a function of the spatial and
mass resolutions. New simulations are underway to address this issue. Presumably
when the numerical resolution becomes higher, the inner parts of subhaloes
are resolved better into higher-density regions that can survive
tidal disruptions longer. However, 
the survival of substructures may be linked to another small-scale
problem of CDM: the central density profiles of low-surface
brightness galaxies (usually with circular velocities
of $\la 100\kms$) seem to be too concentrated compared with observed
galaxies (e.g., Bolatto et al. 2002; Weldrake et al. 2003;
see, however, Swaters et al. 2003). Therefore, if we put in observed mass
profiles, substructures may actually be more
easily destroyed by tidal forces due to their lower central
concentrations. There is another effect that
makes the survival of substructures in the central part more
difficult. In collisionless numerical simulations, the density profile
can be approximated by an NFW profile; the density scales as 
$\propto r^{-1}$ out to $\sim 0.1\rv$ ($\sim 25$ kpc)
for typical galactic-sized haloes. However, the observed velocity
dispersion is nearly constant in the inner part
implying $\rho \propto r^{-2}$, i.e., the density in real galaxies
rises more quickly as the radius decreases. Hence substructures will be disrupted more
easily if they come close to the center and dynamical frictions dragging
them into the center would also be larger in realistically simulated
galaxies.

Our simulations resolve substructure masses from $10^8
M_\odot$ to $10^{11} M_\odot$ for a $10^{12}M_\odot$ parent halo.
However, according to Metcalf et al. (2003), the
required substructure mass is in the range of $10^4M_\odot-10^8 M_\odot$ in
the case of 2237+0305. If the mass spectrum of substructures $n(M)dM \propto
M^{-1.8} dM$ extends all the way down
to $10^4 M_\odot$, one can estimate that the fraction of mass
in substructures with $10^4M_\odot<M <10^8 M_\odot$ is about a factor
of 3 smaller than that in substructures with $10^8M_\odot < M <M^{11}M_\odot$,
i.e., the mass fraction in substructures in the
range required by Metcalf et al. (2003) will be even lower than our
predicted value. This conclusion depends on the
mass spectrum. However, for any power-law spectrum with 
a slope shallower than $-2$, the mass fraction in substructures 
with $10^4 M_\odot<M <10^8 M_\odot$ is likely 
smaller than that in higher-mass substructures identified in
numerical simulations.

There is another effect that may reduce the utilizable mass in
substructures even further: the need for compactness, an issue we
already touched upon in \S2 (see Fig. 1).
In order to affect the flux ratios significantly, the
physical size of substructures must be sufficiently
compact (cf. Metcalf et al. 2003). The most efficient substructures should be of
the same order of the image separation of the pairs or 
triples. For the five CLASS quadrupole lenses that have known lens and source
redshifts (see \S4), the closest pairs have projected separations
from $0.6h^{-1}$kpc (for B1422+231) to $2.3h^{-1}$kpc (for MG0414+0534).
At redshift of 0.5, dark matter haloes with $M \la 4\times 10^6M_\odot$
will have $\rv \sim 1h^{-1}$kpc, and so
they will be efficient in causing flux anomalies if they are
located between close pairs or triples. For larger masses, the
effect is reduced. We estimate the reduction by assuming that only the mass enclosed within
$1h^{-1}$kpc will affect the flux ratios. We adopt a mass spectrum for
dark matter haloes of $n(M) dM \propto M^{-1.8}$ for $10^4M_\odot < M < 10^{11} M_\odot$.
Each halo is described by an NFW profile with a concentration
parameter given by $c\approx 11(M/10^{12}h^{-1})^{0.15}$ (e.g.,
Zhao et al. 2003). Note that the substructures in the central parts 
are tidally truncated and so cannot be fitted well by NFW profiles, but
as most substructures have $r>0.1\rv$ in spherical
radius, the effect of tidal truncation may be modest. We find that
the mass in substructures that
can efficiently cause flux anomalies is reduced by an additional factor of 
five compared with the total mass in all substructures --
most substructures which are in the outer parts are too extended to
cause flux anomalies efficiently. Our rough estimate
shows that the compactness requirement is an issue that
needs to be addressed more carefully. The two additional effects noted
can reduce the likely mass fraction in substructures having the required 
masses and sizes to as low as $\sim 0.03\%$, uncomfortably low compared
with the observational requirement.

  Progress can be made from both the observational and theoretical fronts
to reduce the uncertainties. Observationally, in the radio, the effect
of scattering can be studied with observations at high frequency where
it is expected to be 
unimportant. In the infrared, it would be interesting to have more observations
with integral field spectroscopy similar to that
reported by Metcalf et al. (2003). This method offers a way to 
separate stellar microlensing from substructure milli-lensing. More
astrometric signatures of substructures will be important as well (see \S2).
Theoretically, higher-resolution simulations 
are needed and are already under-way. If future observations and numerical simulations
still indicate a discrepancy between lensing requirements and CDM
predictions, then alternatives must be sought.
One possibility is that these substructures are
massive black holes with $M \sim 10^5-10^6 M_\odot$ (Lacey \& Ostriker 1985; Xu
\& Ostriker 1994), which satisfy the mass and compactness requirements.
We require only a few percent of the surface density
in the substructures, so the density parameter in these
black holes is only $\fsub\Omega_0 \sim 0.012 (\fsub/0.04)$,
which is about $30\%(\fsub/0.04)$ of the baryon density in the
universe. These massive black holes will have other
observable signatures (e.g., Wambsganss \& Paczy\'nski 1992; Tremaine
\& Ostriker 1999) and can be further tested.

\acknowledgments 

We thank I. Browne, H. J. Mo, T. York and S. White for helpful discussions, X. Kang for
identifying subhalos and V. Springel for providing us with his {\tt SUBFIND} code.
YPJ is supported in part by NKBRSF (G19990754) and by NSFC.

{}

\begin{figure}
\epsscale{0.8}
\plotone{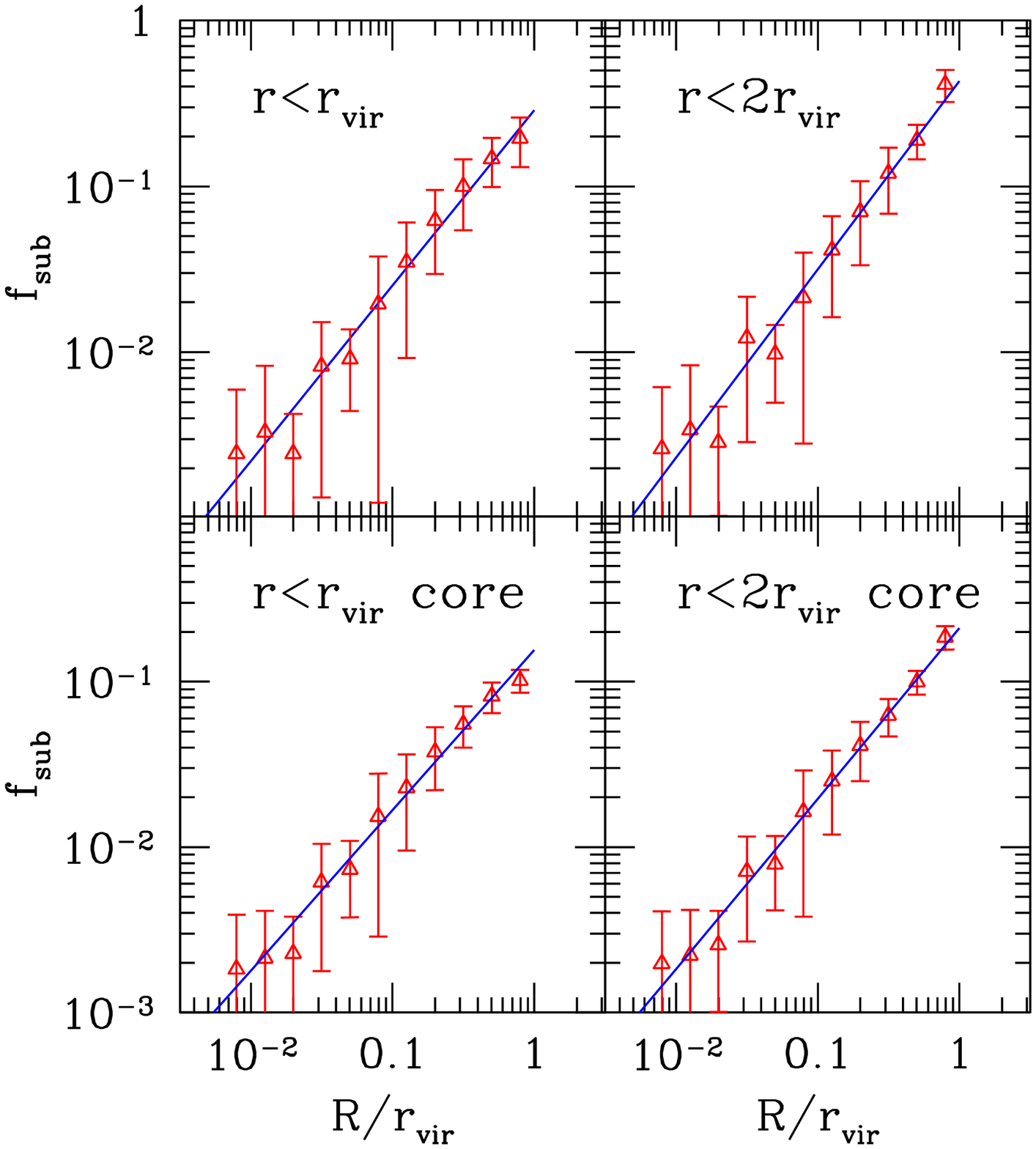}
\caption{
Predicted fractions of dark matter surface mass density in
substructures, $\fsub$,
as a function of the projected radius, $R$, in units of the virial radius, $\rv$.
The lensed images typically have $R/\rv =1\%-3\%$.
The left panels show $\fsub$ as a function of $R$ for 
all substructures within the spherical radius $r<\rv$ while the right
panels include all substructures with $r<2\rv$. 
In the lower panels we only include the mass
within $0.025\rv$ for each subhalo. The solid line in each panel
indicates an unweighted least-square fit ($\log \fsub=\log A + \log
R/\rv$, i.e., $\fsub=A R/\rv$).
The $A$ values are, 0.25, 0.31, 0.20 and 0.17, clockwise from
the top left to the bottom right panels, respectively.
}
\label{fig:fig1}
\end{figure}


\begin{references}
\reference{} Biggs, A. et al. 2003, \mnras, submitted
\reference{} Bode, P, Ostriker J. P. 2003, \apjs, 145, 1
\reference{} Bolatto, A. D., Simon, J. D., Leroy, A., Blitz, L. 2002,
	\apj, 565, 238
\reference{} Bradac, M., Schneider, P., Steinmetz, M., 
	Lombardi, M., King, L. J., Porcas R. 2002, \aap, 388, 373
\reference{} Browne, I.W.A., et al. 2003, \mnras, 341, 13
\reference{} Chen, J., Kravtsov, A.V., Keeton, C. R. 2003, \apj, 592, 24
\reference{} Chiba, M. 2002, ApJ, 565, 17
\reference{} Dalal, N., \& Kochanek, C. S. 2002, \apj, 572, 25
\reference{} Evans, N. W., \& Witt, H. J. 2003, 
	\mnras, 345, 1351
\reference{} Fassnacht, C. D. et al. 1999, \aj, 117, 658
\reference{} Ghigna, S., Moore, B. Governato, F., Lake, G., Quinn, T.,
	Stadel, J. 2000, \apj, 544, 616
\reference{} Hewitt, J. N. et al. 1992, \aj, 104, 968
\reference{} Ibata, R., Irwin, M., Lewis, G. F., Stolte, A. 2001, 
      ApJ, 547, L133
\reference{} Jackson, N. J. et al. 1998, \mnras, 296, 483
\reference{} Jing, Y. P. \& Suto Y. 2000, ApJ, 529, L69
\reference{} Jing, Y. P. \& Suto Y. 2002, ApJ, 574, 538
\reference{} Lacey, C. G., \& Ostriker, J. P. 1985, \apj, 299, 633
\reference{} Kauffmann, G., White, S. D. M., Guiderdoni, B.  1993,
	\mnras, 264, 201
\reference{} Keeton, C. R. 2001, \apj, 561, 46
\reference{} Keeton, C. R. 2003, \apj, 584, 664
\reference{} Keeton, C. R., Kochanek, C. S., Seljak, U. 1997, ApJ, 482, 604
\reference{} Klypin, A., Kravtsov A. V., Valenzuela O. 1999, \apj, 522,
	82
\reference{} Kochanek, C.S. 1991, ApJ, 373, 354
\reference{} Kochanek, C.S. et al. 2000, \apj, 543, 131
\reference{} Kochanek, C.S., \& Dalal, N. 2002, preprint (astro-ph/0212274)
\reference{} Kochanek, C.S., \& Dalal, N. 2003, preprint (astro-ph/0302036)
\reference{} Koopmans, L. V. E., Fassnacht, C. D. 1999, \aj, 123, 627
\reference{} Koopmans, L. V. E. et al. 2002, \apj, 334, 39
\reference{} Koopmans, L. V. E. et al. 2003, \apj, 595, 712
\reference{} Mao, S., Schneider, P. 1998, \mnras, 295, 587
\reference{} Marlow, D. et al. 1999, \aj, 118, 654
\reference{} Metcalf, R. B. 2002, 580, 696
\reference{} Metcalf, R. B., \& Madau, P. 2001, \apj, 563, 9
\reference{} Metcalf, R. B., \& Zhao, H.S. 2002, \apj, 567, L5
\reference{} Metcalf, R. B., Moustakas, L. A., Bunker, A. J., Parry,
        I. R. 2003, astro-ph/0309738
\reference{} Mo, H. J., Mao, S., White, S. D. M. 1998, \mnras, 295, 319
\reference{} Moore, B., Ghigna, S., Governato, F. et al. 1999, \apj,
	524, L19
\reference{} Moustakas, L.A., \& Metcalf R. B. 2003, \mnras, 339, 607
\reference{} Myers, S. T. et al. 1995, \apj, 447, L5
\reference{} Myers, S. T. et al. 2003, \mnras, 341, 1
\reference{} Patnaik, A. R. et al. 1992, \mnras, 259, 1P
\reference{} Patnaik, A. R., \& Narasimha, 2001 \mnras, 326, 1403
\reference{} Phillips, P. M.  et al. 2000, \mnras, 319, L7
\reference{} Romanowsky, A. J. et al. 2003, astro-ph/0308518
\reference{} Navarro, J. F., Frenk, C. S., White, S. D. M. 1997,
	\apj, 490, 493
\reference{} Schechter, P. L., \& Wambsganss J. 2002, \apj, 580, 685
\reference{} Schechter, P. L., \& Wambsganss J. 2003, astro-ph/0309163
\reference{} Springel, V., White, S.~D.~M., Tormen, G., \& Kauffmann, G.\ 2001, \mnras, 
328, 726 
\reference{} Spergel, D.N.S. et al. 2003, \apjs, 148, 175
\reference{} Stoehr, F., White, S. D. M., Tormen, G., Springel, V. 2002,
         \mnras, 335, L84
\reference{} Swaters, R. A., Madore, B. F., van den Bosch, F. C.,
	Balcells, M. 2003, \apj, 583, 732
\reference{} Tremaine, S., \& Ostriker, J. P. 1999, \mnras, 306, 602
\reference{} Wambsganss, J., \& Paczy\'nski, B. 1992, \apj, 397, L1
\reference{} Weldrake, D. T. F., de Blok, W. J. G., Walter, F. 2003, 
	\mnras, 340, 12
\reference{} Xu, G., \& Ostriker, J. P. 1994, \apj, 437, 184
\reference{} Yanny, B. et al. 2003, \apj, 588, 824
\reference{} Zhao, D. H., Jing, Y. P., Mo, H. J., \& Boerner G. 2003
\apj, 597, L9
\end{references}
\end{document}